\documentclass{iau}
\usepackage{graphicx,natbib}
 
\newcommand{\apjl}{ApJ}           
\newcommand{\mnras}{MNRAS}       

\newcommand{\aap}{A\&A}
\newcommand{\araa}{ARA\&A}

\title{Spectral synthesis of stellar populations in the 3D era: The CALIFA experience}

\author[CALIFA collaboration]{R. Cid Fernandes$^1$, E. A. D. Lacerda$^1$, R. M. Gonz\'alez Delgado$^2$, ~~~~ 
N. Vale Asari$^1$, R. Garc\'\i a-Benito$^2$, E. P\'erez$^2$, A. L. de Amorim$^1$, ~~~ C. Cortijo-Ferrero$^1$,   R. L\'opez Fern\'andez$^2$,  S. F. S\'anchez$^{2, 3}$, \and the CALIFA collaboration}

\affiliation{
$^1$Departamento de F\'\i sica, Universidade Federal de Santa Catarina, 
Florian\'opolis, SC, Brazil \\
email: {\tt cid@astro.ufsc,br}\\
$^2$Instituto de Astrof\'\i sica de Andaluc\'\i a (CSIC), 
Granada, Spain  \\
$^3$ Instituto de Astronom\'\i a, Universidad Nacional Auton\'oma
de Mexico, 
D.F., M\'exico}
\pubyear{2014}
\volume{309}
\jname{Galaxies in 3D across the Universe}
\editors{B. L. Ziegler, F. Combes, H. Dannerbauer, M. Verdugo, eds.}

\begin{document}

\maketitle

\begin{abstract} Methods to recover the fossil record of galaxy evolution encoded in their optical spectra have been instrumental in processing the avalanche of data from mega-surveys along the last decade, effectively transforming observed spectra onto a long and rich list of physical properties: from stellar masses and mean ages to full star formation histories. This promoted progress in our understanding of galaxies as a whole. Yet, the lack of spatial resolution introduces undesirable aperture effects, and hampers advances on the internal physics of galaxies. This is now changing with 3D surveys. The mapping of stellar populations in data-cubes allows us to figure what comes from where,  unscrambling information previously available only in integrated form.
This contribution uses our {\sc starlight}-based analysis of 300 CALIFA galaxies to illustrate the power of spectral synthesis applied to data-cubes. The selected results highlighted here include: (a) The  evolution of the mass-metallicity and mass-density-metallicity relations, as traced by the mean stellar metallicity. (b) A comparison of star formation rates obtained from H$\alpha$ to those derived from full spectral fits. (c) The relation between star formation rate and dust optical depth within galaxies, which turns out to mimic the Schmidt-Kennicutt law. (d) PCA 
tomography experiments.

\end{abstract}

\firstsection

\section{Introduction}

So much has changed so quickly in the way we study galaxies that an astronomer suddenly transported from a conference in the 1990s to a symposium like this one would be thoroughly stunned with things we now take for granted, like counting galaxies by the thousands, and unashamedly quoting their stellar masses, star formation rates and mean stellar ages as if these properties were trivially derived. This revolution happened due to the confluence of major advances in two fronts. First, mega-surveys like the SDSS came into existence. Secondly, models for the spectra of stellar populations with an appropriate quality and resolution to analyze actual galaxy spectra finally became available. This gave new life to old spectral synthesis methods to dig the fossil record of galaxy evolution out of their observed spectra, i.e., to transform $F_\lambda$ into physical properties such as stellar mass, mean age, metallicity, extinction, and the whole star formation history (SFH).

Naturally, there are caveats. On the modeling side---skipping technical aspects and the semi-philosophical issue of whether to use index-based or  full $\lambda$-by-$\lambda$ spectral fitting methods---uncertainties in the evolutionary synthesis ingredients are currently an important limiting factor (e.g., Chen et al.\ 2010; Cid Fernandes et al.\ 2014). On the observational side, though much has been learned from the application of spectral synthesis to SDSS galaxies, the lack of spatial resolution severely limits interpretation of the results. Indeed, not knowing which photons come from where forces us to treat galaxies as point sources, mixing clearly distinct sub-components (bulge, disk, arms, bars) and their different stellar populations. 

Here is where Integral Field Spectroscopy (IFS) surveys come to our rescue (see S\'anchez contribution), combining the morphological information of images with the diagnostic power of spectroscopy. Besides producing  maps of everything we previously had only for entire galaxies, these surveys will be instrumental in reverse engineering the meaning of global properties derived from spatially unresolved data like the SDSS and high-$z$ targets.

We have been exploring the information on stellar populations within CALIFA datacubes by means of the spectral synthesis code {\sc starlight}. In short, after some basic preprocessing steps the datacube is spatially binned into Voronoi zones whenever necessary to reach $S/N \ge 20$, and all zone spectra are extracted and fitted with {\sc starlight}. The results are then packed into fits files and analyzed  with  {\sc p}y{\sc casso}\footnote{Python Califa Starlight Synthesis Organizer}, a very handy tool to post-process and analyze  {\sc starlight} results in 3D. P\'erez et al.\ (2013) and Gonz\'alez Delgado et al.\ (2014a,b) present our first scientific results, while technical aspects are  described in Cid Fernandes et al.\ (2013, 2014). Instead of reviewing this already published work, we dedicate these few pages to presenting some yet unpublished (and unpolished) results which further illustrate the richness of the manifold of physical properties derived from the application of stellar populations diagnostic tools to IFS surveys.

\section{Chemical evolution}

Recently, Gonz\'alez Delgado et al.\ (2014b) analyzed the {\sc starlight}-derived mean {\it stellar} metallicities ($Z_\star$) of 300 CALIFA data cubes. Metallicity  was shown to correlate both with the stellar mass ($M_\star$) and the surface mass density ($\mu_\star$). $Z_\star$ is more closely related to $M_\star$ in spheroids, while in disks it is $\mu_\star$ who seems to govern the chemical evolution. Thus, as we had previously found for mean stellar ages, the balance between local ($\mu_\star$-driven) and global ($M_\star$-driven) processes affecting $Z_\star$ varies with the location within a galaxy (see also Gonz\'alez Delgado's contribution in this same volume).

\begin{figure}
\centering
\includegraphics[width=0.495\columnwidth]{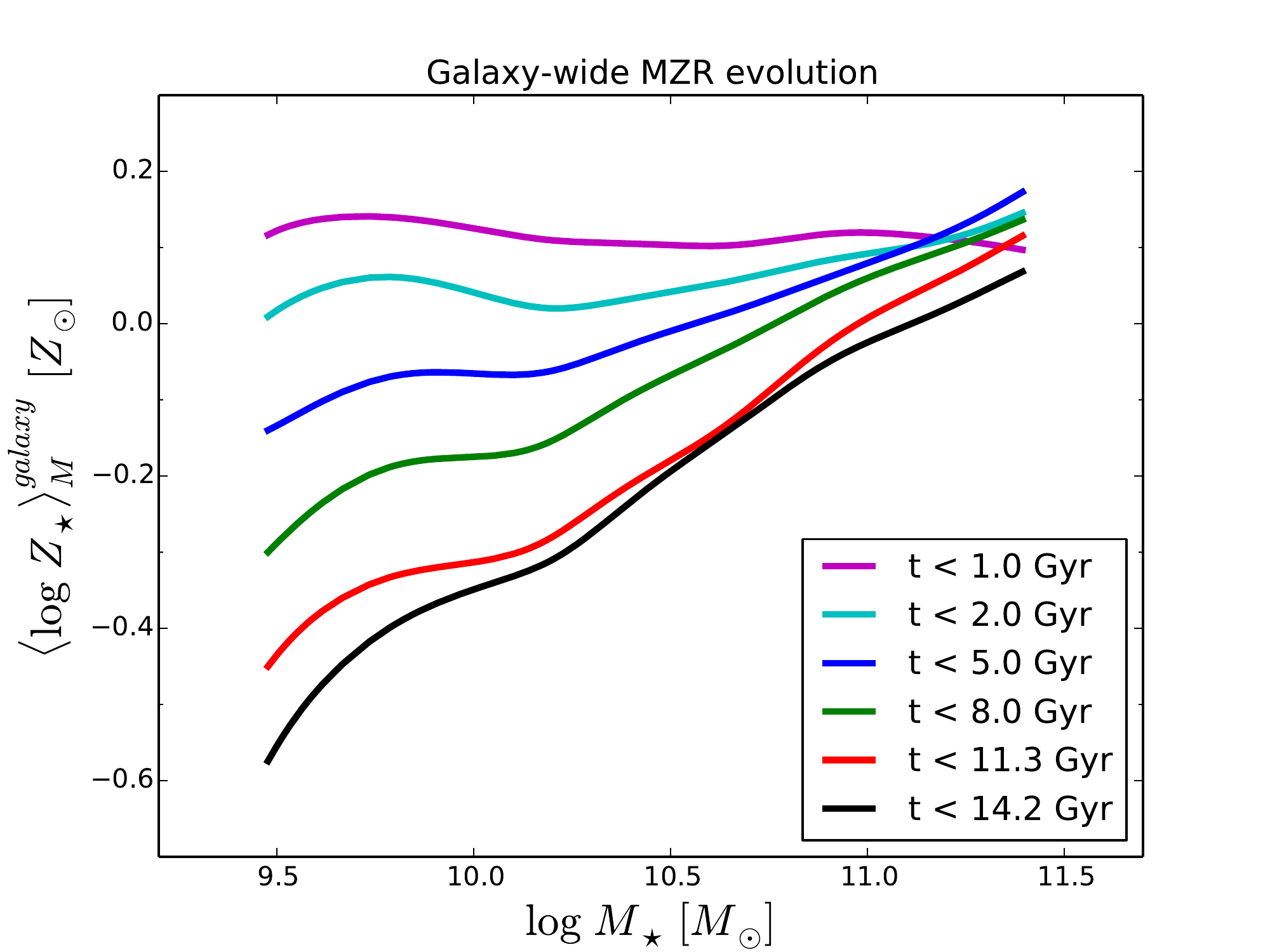}
\includegraphics[width=0.495\columnwidth]{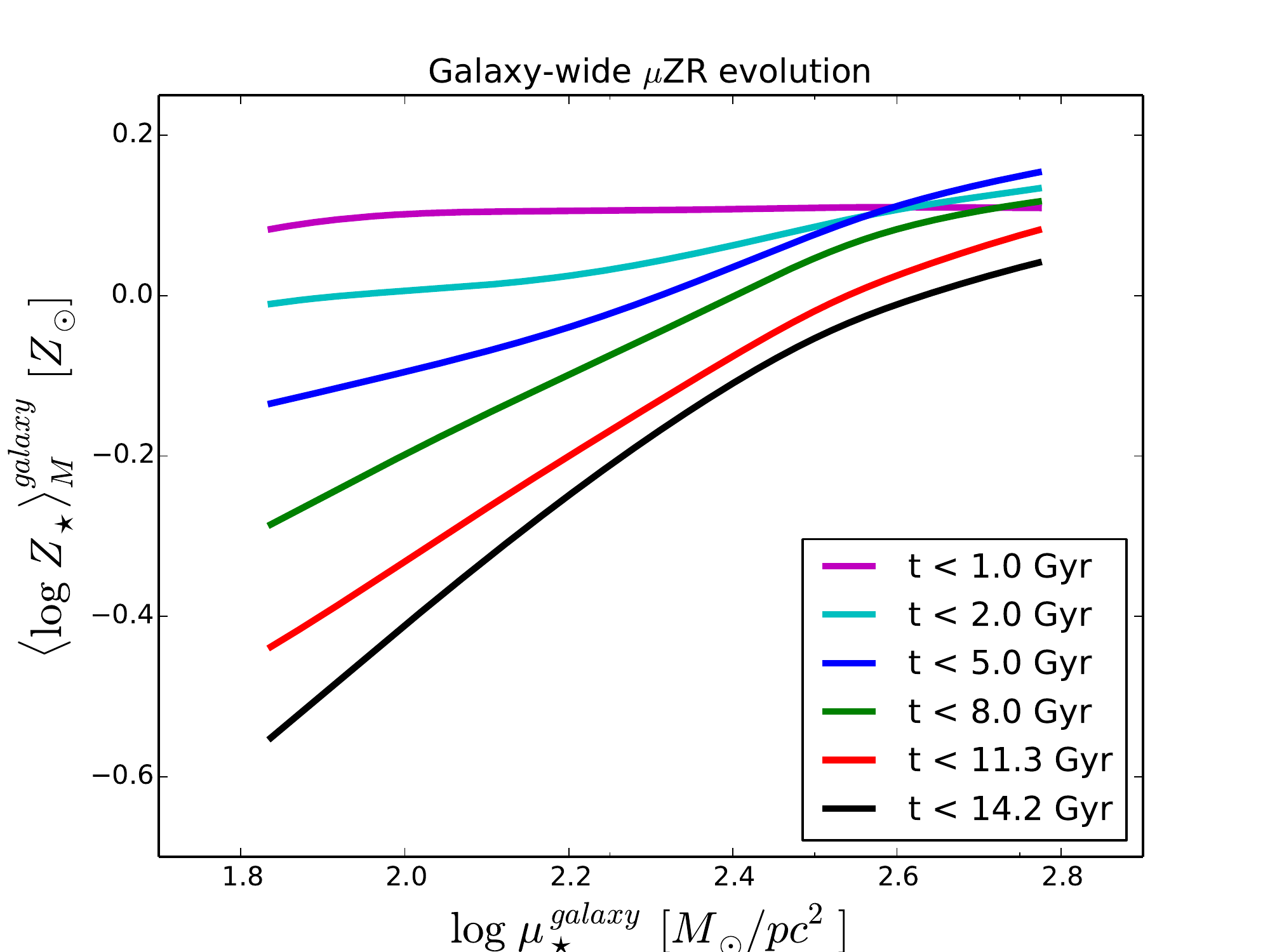}
\caption{ 
Evolution of the galaxy-wide $M_\star$  and $\mu_\star$ versus {\it stellar} metallicity relations. Each line corresponds to the  average $Z_\star(<t)$ curve obtained from 300 galaxies ($\sim 250$k spectra). $Z_\star(<t)$ values are computed including only stellar populations younger than $t = 1, 2, \ldots 14$ Gyr (from top to bottom), as derived from the {\sc starlight} spectral decomposition.
}
\label{fig:ChemicalEvol}
\end{figure}

Another finding reported in that paper is that the $Z_\star$ values obtained considering only stars younger than 2 Gyr are  higher than those for the whole population (1 Myr $< t < $ 14 Gyr). Although this is expected in terms of chemical evolution, $Z_\star$ is a non-trivial property to derive in composite stellar populations, and deriving its time dependence is even harder. In Fig.\  \ref{fig:ChemicalEvol} we look in more detail at this remarkable result. Its left panel shows the $M_\star$-$Z_\star$ relation (MZR), breaking $Z_\star(<t)$ into six age ranges, from $t < 14$ Gyr (i.e., all stars) to $t < 1$ Gyr. Similarly, the right panel shows the evolution of the $\mu_\star$-$Z_\star$ relation ($\mu$ZR). The curves are based on 300 galaxies (but 253418 spectra!), whose galaxy-wide average $Z_\star(<t)$ are binned along the horizontal axis and smoothed for cosmetic purposes.

Both plots show that $Z_\star$ increases steadily as galaxies age. The pace of chemical evolution varies with both $M_\star$ and $\mu_\star$, producing the systematic flattening of the MZR and $\mu$ZR with time. In fact, stars of all ages have similar metallicities in the most massive (also the densest) galaxies, as these systems essentially completed their star formation and chemical evolution very long ago. Going down the $M_\star$ and $\mu_\star$ scales one finds slower and slower evolution. The SFHs of these galaxies are also slower, in the sense that their stellar mass was formed over a longer time-span.

The point to highlight here is that $Z_\star$ and its time evolution behave amazingly well, and for galaxies of all types. Studies of the MZR are usually based on nebular properties. Though useful, $Z_{neb}$ can only be applied when the emission lines are powered by young stars, thus excluding regions where BPT-like diagnostic diagrams indicate the contribution of other ionizing agents (AGN, shocks, old stars). Furthermore, it is obviously impossible to trace the evolution of $Z_{neb}$ for any single galaxy.  The availability of reliable estimates of the stellar metallicity thus offers an independent (and in several ways more informative)  way to address issues related to the chemical evolution of galaxies.

\section{Star formation Rates: Nebular $\times$ stellar tracers}

By construction, estimates of the recent star formation rate (SFR)  involve the hypothesis that SFR $=$ constant over some time interval $T$. In the case of the H$\alpha$ luminosity, one has $L_{H\alpha} \propto \int_0^{ ``\infty"} {\rm SFR}(t)  q_H(t) dt \sim {\rm SFR} \int_0^T  q_H(t) dt$, where $q_H(t)$ is the rate of ionizing photons per unit initial mass of a burst of age $t$. Because $q_H(t)$ drops precipitously after  $\sim 10^7$ yr,  its integral converges on a time-span $T$ of this same order.\footnote{Other tracers like the UV luminosity or the dust-reprocessed far-IR luminosity follow the same logic, but the corresponding time-integrals require much longer times to converge.} Assuming  SFR$(t<T) \sim$ constant seems plausible when dealing with whole (or large chunks of) galaxies, since in this regime the data  average over many regions formed in the last $T$ years (e.g., a collection of HII regions of different ages and stellar masses). Doing the same for a single spaxel is at least dangerous, if not wrong, as one quickly realizes by imagining the case of a spaxel containing a single HII region, understood as an instantaneous burst where $L_{H\alpha}$ does not measure a rate (see also Calzetti's contribution).

\begin{figure}
\centering
\includegraphics[width=1\columnwidth]{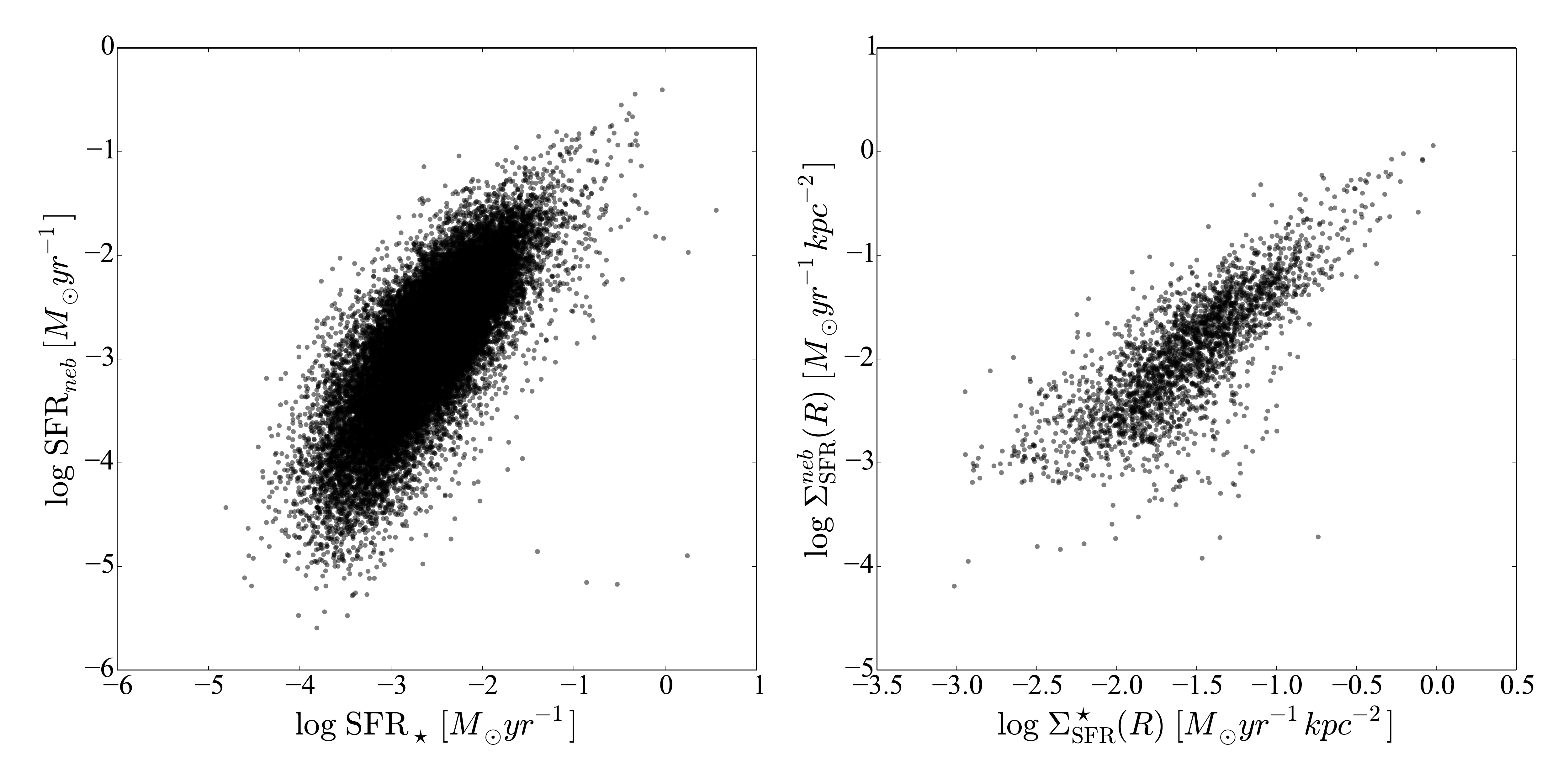}
\caption{ 
{\it Left:} Comparison of the SFRs estimated from H$\alpha$ with those obtained from the spectral fits of the stellar continuum for $\sim 40$k CALIFA spectra. {\it Right:} As in the left, but now comparing  SFR surface densities (thus eliminating distance-induced correlations) for 20 radial bins per galaxy, hence averaging over HII regions, as implicitly required by the SFR(H$\alpha$) method. The {\sc starlight}-based SFRs were computed for a time-scale of 25 Myr.
}
\label{fig:SFRs}
\end{figure}

In contrast, {\sc starlight} assumes nothing about the SFH. Its more general and flexible non-parametric description of the SFH is thus in principle more suitable to derive SFRs, and these can be computed over any desired time scale. The caveat is that, unlike for ionizing photons, the optical continuum contains a non-trivial mixture of populations of all ages, and isolating the contribution due to a specific sub-population is an inevitably uncertain and degenerate process. Still, Asari et al.\ (2007) showed that the H$\alpha$ and {\sc starlight}-based SFRs agree remarkably well in SDSS star-forming galaxies, so we repeat this comparison for CALIFA data.

In Fig.\ \ref{fig:SFRs} we compare the H$\alpha$ and {\sc starlight}-based SFRs for 40532 zones in 300 CALIFA galaxies, selected to belong to the star-forming wing in the [OIII]/H$\beta$ vs.\ [NII]/H$\alpha$ diagram. Despite all potential caveats pointed out above, the correlation is excellent. In the right panel we refine the analysis by comparing surface densities averaged over radial rings, thus mitigating stochastic effects and better complying with the SFR $\sim$ const.\ hypothesis implicit in the H$\alpha$ method. The correlation is even better, and in fact cleaner.

One can look at this result as reinforcing the confidence on the spectral fits with {\sc starlight}.  One can also look at it in closer detail to map  differences between nebular and stellar estimates of SFRs, and to investigate plausible causes such as differential extinction and IMF. Also, now that we have a SFR indicator independent of (but consistent with) emission lines, we can use it across the board, even when SFR(H$\alpha$) does not apply, as in AGN dominated regions. We leave this as a small demonstration of the power (and promise) of the combination of stellar and nebular diagnostics.

\section{Schmidt-Kennicutt like relations}

\begin{figure}
\centering
\includegraphics[width=0.49\columnwidth]{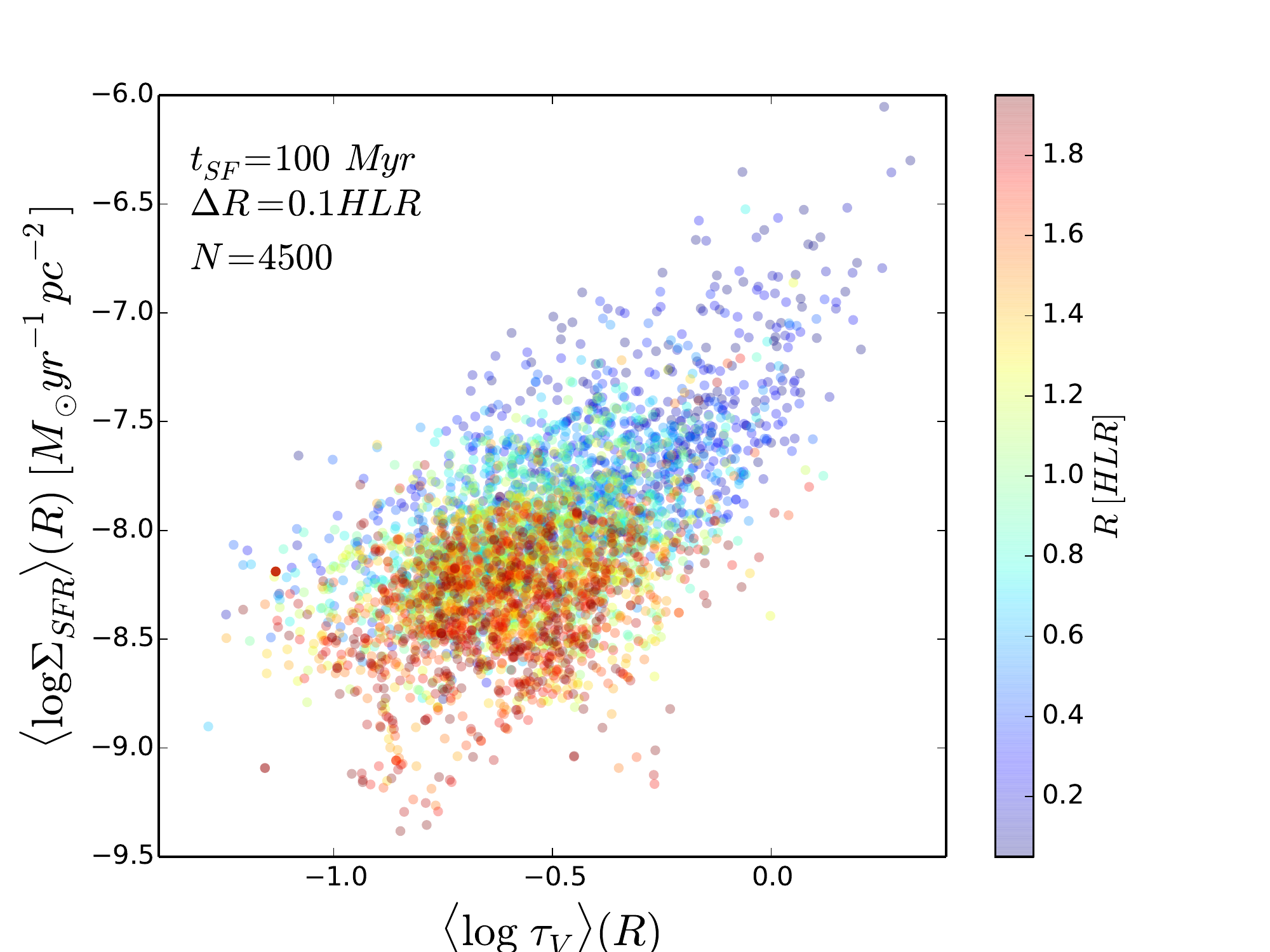}
\includegraphics[width=0.49\columnwidth]{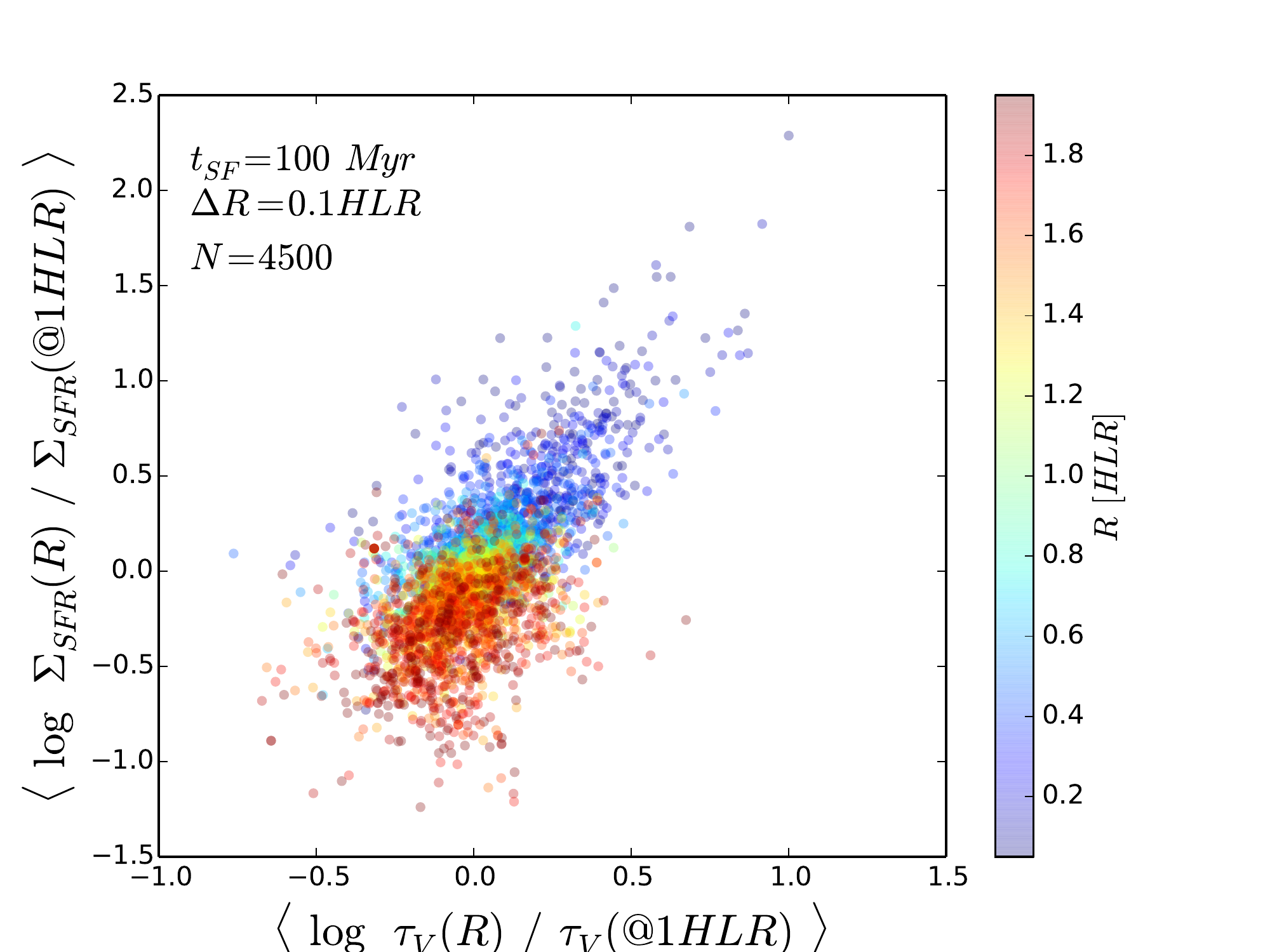}
\caption{{\it Left:} Relation between SFR surface density and dust optical depth, both derived from the spectral synthesis of 300 CALIFA galaxies. Each point corresponds to radial averages of $\Sigma_{SFR}$ and $\tau_V$  for an (inclination corrected) radial bin of width $\Delta R = 0.1$ Half Light Radius (HLR), color coded to reflect $R$.
{\it Right:} As the left panel, but scaling the $\Sigma_{SFR}$ and $\tau_V$ values of each galaxy by the corresponding values at $R = 1$ HLR.
Barring a dust-to-gast conversion factor, these relations are analogous to the  Schmidt-Kennicutt relation between $\Sigma_{SFR}$ and $\Sigma_{gas}$.
}
\label{fig:SK}
\end{figure}

As any optical survey, CALIFA looks at the stellar, nebular, and dust content of galaxies. The missing piece of information is gas, the fuel for star formation and galaxy evolution. 
Molecular and atomic maps of CALIFA galaxies will  tie in nicely with the information already at hand, allowing spatially resolved studies of the relation between  gaseous content, SFR and chemical abundance  for a varied and representative sample.

Meanwhile, let us explore dust as proxy for gas. Specifically, we correlate the V-band dust optical depth ($\tau_V$) with the SFR surface density ($\Sigma_{SFR}$). Both quantities are derived entirely from the {\sc starlight} spectral fits. This allows a more comprehensive view than one would obtain by restricting the analysis to spaxels whose emission lines are both (a) consistent with star-formation in BPT-like diagrams, and (b) strong enough to be reliably measured.  Since  $\tau_V$ ultimately represents the dust column density $\Sigma_{dust}$, which in turn presumably traces $\Sigma_{gas}$, to first order (and ignoring metallicity dependence in the dust-to-gas conversion factor) one expects $\Sigma_{SFR}(\tau_V)$ to behave as a Schmidt-Kennicutt  $\Sigma_{SFR}(\Sigma_{gas})$  relation.

This expectation is borne out in Fig.\ \ref{fig:SK}, which shows the relation between $\Sigma_{SFR}$ and $\tau_V$ we obtain with CALIFA. The plot contains 4500 points, each one with average measurements performed over radial bins of width $=  0.1$ Half Light Radius (HLR). Both $\Sigma_{SFR}$ and $\tau_V$ increase towards the nucleus, as one infers from the color change from the upper-right to bottom-left of the plot.  Despite the scatter, the relation is visibly super-linear, with a logarithmic slope in the neighborhood of 1.5, close to what is found with direct measurements of $\Sigma_{gas}$ (Kennicutt \& Evans 2012 and references therein). The right panel in Fig.\ \ref{fig:SK} rescales the $\Sigma_{SFR}$ and $\tau_V$ values of each galaxy to the corresponding values at $R = 1$ HLR. This rescaling trick, which ultimately expresses results for each galaxy in its own natural units, significantly reduces the scatter, probably due to the minimization of ``second-parameter" effects.

Again, we show these plots as a teaser of what is nowadays doable thanks to the combination of spectral synthesis tools and 3D spectroscopic surveys. Empirical relations such as the ones in Fig.\ \ref{fig:SK} can be worked from different angles. One can for instance use them to verify the existence of a SK-like relation within galaxies and investigate how the star-formation ``efficiency" varies with, say, morphological type or mass-density. Alternatively, one may postulate a SK law and use $\Sigma_{SFR}$ to derive $\Sigma_{gas}$, which in turn would be useful to study (i) $\Sigma_{dust}/\Sigma_{gas}$ dust-to-gas ratios, (ii) gas fractions and their relation to stellar and nebular metallicities, (iii) etc. Plenty of work ahead!

\section{PCA tomography}

Finally, we briefly mention an entirely different approach to explore data-cubes. The PCA tomography technique devised by Steiner et al.\ (2009) identifies correlations within data-cubes in a purely mathematical way, free of astrophysical pre-conceptions. Reverse engineering the meaning of the results is not always a simple task, although  spectacular results achieved with Gemini data-cubes reveal the potential of this technique. 

\begin{figure}
\centering
\includegraphics[width=0.99\columnwidth]{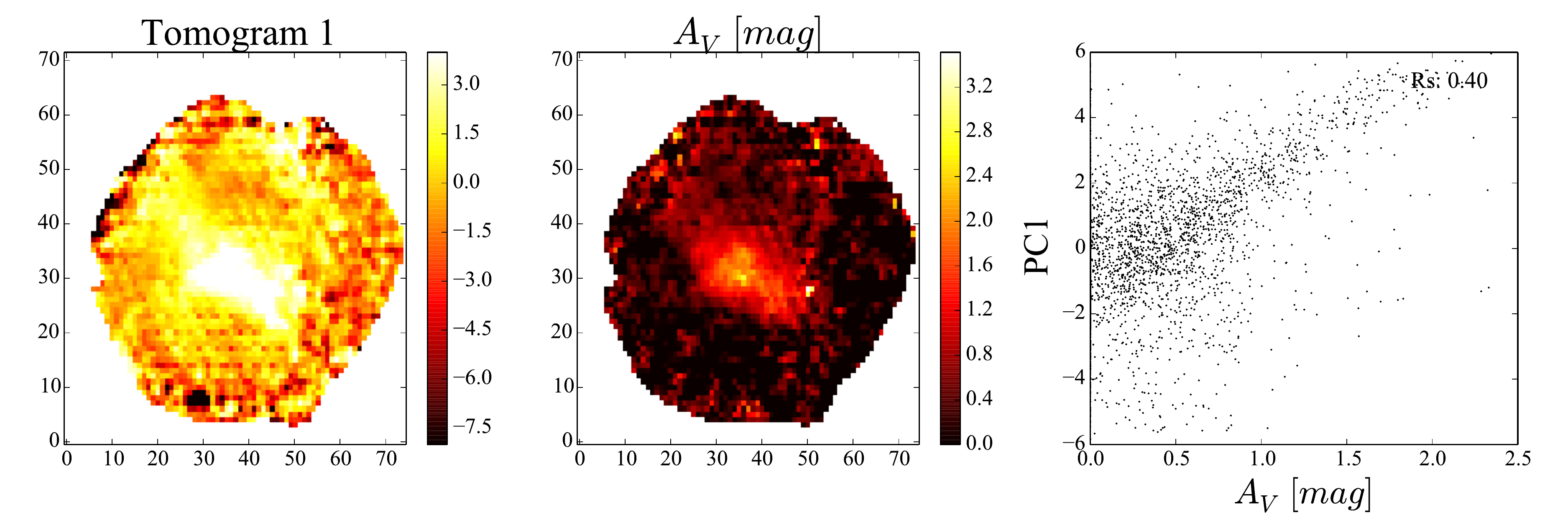}
\caption{PCA tomography experiments with CALIFA data-cubes (see text for details).
}
\label{fig:PCA}
\end{figure}

Fig.\ \ref{fig:PCA} gives a taste of what one finds for CALIFA data after masking emission lines and rescaling fluxes to account for the large dynamic range within the cube---PCA is highly sensitive to pre-processing steps. The left panel shows the image (tomogram) of the first PC in the data-cube of Arp 220. The central lane of high PC1 reflects the region of large extinction, as confirmed by its clean correspondence with the {\sc starlight}-derived $A_V$ map in the central panel and the PC1 vs.\  $A_V$ plot on the right. Considering the few galaxies analyzed so far, we generally find that the first 2 or 3 PCs generally correlate well with some stellar population property. The velocity field, in particular, is invariably recognized in at least one PC. From our point of view this is actually a handicap of the method, as variance is waisted in mapping a property ($v_\star$) which can easily be measured by more conventional  means, so ways to circumvent this drawback are being implemented. We also note that relating PCs to previously known properties is an instructive exercise, although it kind of defeats the very purpose of processing the data through a physics-free algorithm, and letting the tomograms and eigenspectra hint at the underlying phenomena. In any case, there is still plenty more to be examined. Emission line and combined stellar plus nebular spectral cubes are still being PCA'ed, and one hopes that this technique helps unveiling the interplay between stars, gas and dust. Last but not least, PCA and other techniques may be used as noise filters and/or to identify subtle instrumental effects, so there is plenty to be done along this line too.

\section*{Acknowledgements}
\noindent
This contribution is based on data obtained by the CALIFA survey (http://califa.caha.es), funded by the Spanish MINECO grants ICTS-2009-10, AYA2010-15081, and the CAHA operated jointly by the Max-Planck IfA and the IAA (CSIC). The CALIFA Collaboration thanks the Calar Alto staff for the dedication to this project. Support from CNPq (Brazil) through grants  300881/2010-0 and 401452/2012-3 is duly acknowledged.

\end{document}